\documentclass[12pt,preprint]{aastex} 
\usepackage{epsfig}
\usepackage{amsmath} 
\usepackage{rotating}

\begin{document}
\title{\Large\bf Design and Performance of a Digital Phase Lock Loop for Gunn Oscillators}
\smallskip
\author{Todd R. Hunter\altaffilmark{1}, Robert Kimberk\altaffilmark{2}, Patrick Steve Leiker\altaffilmark{2}, Cheuk-Yu Edward Tong\altaffilmark{2} \& Robert W. Wilson\altaffilmark{2}}
\email{thunter@nrao.edu}
\altaffiltext{1}{NRAO, 520 Edgemont Rd, Charlottesville, VA, 22903}
\altaffiltext{2}{Harvard-Smithsonian Center for Astrophysics, 60 Garden St., Cambridge, MA, 02138}


\begin{abstract}

The digital phase lock loop described in this paper has been in use on
the Submillimeter Array (SMA) front-end receivers for over a decade
and has been a key element in obtaining all of the high quality images
that have been published from this telescope over the years.  The
technical achievements enabled by these devices include the first
phase closure observations in the 690 GHz band, the first attempts at
band-to-band phase transfer at submillimeter wavelengths, and the
first successful demonstration of interferometry using a fully
photonic millimeter-wave local oscillator. Copies of these devices are
also in place at the Caltech Submillimeter Observatory and the James
Clerk Maxwell Telescope in support of the eSMA project and
submillimeter VLBI experiments.  Additional units of this design were
used by the Princeton Millimeter Interferometer and the Microwave
Anisotropy Telescope.  In total, over three dozen units have been
constructed and used in astronomical studies.  In this paper, we
briefly describe the background theory, design, performance, and
calibration steps, and provide useful testing and repair information.

\end{abstract}

\section{Introduction}

Operating as a multi-element interferometer, the Submillimeter Array
(SMA) located on Mauna Kea, Hawaii enables astronomical imaging with
substantially finer angular resolution than the $10\arcsec$ beams
achievable by typical single-dish submillimeter cameras and receivers
\citep[see e.g.][]{Hunter1996,Kawamura1999}.  Standard radio
interferometers require a phase-locked local oscillator (LO) signal in
order to achieve frequency conversion from the desired high frequency
observing band to a more digitally-accessible intermediate frequency
(IF) band \citep{Thompson2001}.  The digital phase lock loop (PLL)
described in this paper has been in use on the SMA front-end receivers
\citep{Blundell2005} for over a decade and has been a key element in
obtaining all of the high quality, sub-arcsecond images that have been
published from this telescope over the years \citep[see
  e.g.][]{Hunter2008}.  The technical achievements enabled by these
devices include the first phase closure observations in the 690 GHz
band \citep{Hunter2007}, the first phase transfer at submillimeter
frequencies \citep{Hunter2006}, and the first successful demonstration
of interferometry using a fully photonic millimeter-wave local
oscillator \citep{Kimberk2006}. For the latter experiment, a small
modification was made to drive a YIG oscillator rather than the
standard Gunn oscillator described in the following section.

Copies of the PLL devices and their control software are also in place
at the Caltech Submillimeter Observatory (CSO) and the James Clerk
Maxwell Telescope (JCMT) in support of the eSMA project, described by
\citet{Bottinelli2008}. As such, they are also used in current and
future submillimeter VLBI observations
\citep{Doeleman2008,Weintroub2008}.  Additional units of this design
were used by the Princeton Millimeter Interferometer (MINT)
\citep{Doriese2002,Fowler2005} and the Microwave Anisotropy Telescope
collaboration of Princeton and University of Pennsylvania
\citep{Torbet1999} both located in the Atacama Desert of northern
Chile.  In total, over three dozen units have been constructed and
used in astronomical studies.  In this paper, we briefly describe the
background theory, design, performance, and calibration steps, and
provide useful testing and repair information. Most of the content of
this paper can also be found in SMA technical memo
143\footnote{http://www.cfa.harvard.edu/sma/memos/143.pdf}.

\section{Background: Gunn oscillator frequency tuning}

The output frequency $\nu$ of a cavity-tuned Gunn oscillator depends
simultaneously on both the electrical and mechanical tuning.
Specifically, $\nu$ is a function of the bias voltage $V$ and the
cavity length $L$:
\begin{equation}
  \nu = f(V,L).  
\end{equation}
At a specified tuning point $i$, the function $f$ can be approximated
by a linear dependence in each variable independently, leading to
the following differential equation:
\begin{equation}
  \nu = \nu_i + \kappa_{V,i} \Delta V_i + \kappa_{L,i} \Delta L_i,
  ~~{\rm where}~\frac{d\nu}{dV}\biggr|_i = \kappa_{V,i}~~{\rm and}~~
  \frac{d\nu}{dL}\biggr|_i = \kappa_{L,i}~.
\end{equation}
Typically, $\kappa_{V,i}$ is listed in units of MHz~volt$^{-1}$ and is
positive-valued, while $\kappa_{L,i}$ can be interpolated from a
micrometer tuning curve listed in units of mil~GHz$^{-1}$ and is
negative-valued.  With the application of a phase lock loop, $\nu$ can
be held constant in time at $\nu_i$ as it can actively shift the bias
voltage by $\Delta V(t)$ in order to offset temporal changes in the
cavity length $\Delta L(t)$.  Much of $\Delta L$ is due to drifts in the
room temperature.   Thus, a successful PLL must be able to maintain
lock through a sufficient range of tuning voltage to counteract
ambient temperature fluctuations.

\section{Design of a Digital PLL}

\begin{figure}[hbt] 
\hspace{2.5in} Gunn bias output \hspace{1.4in} Harmonic mixer\\
\centerline{\epsfig{file=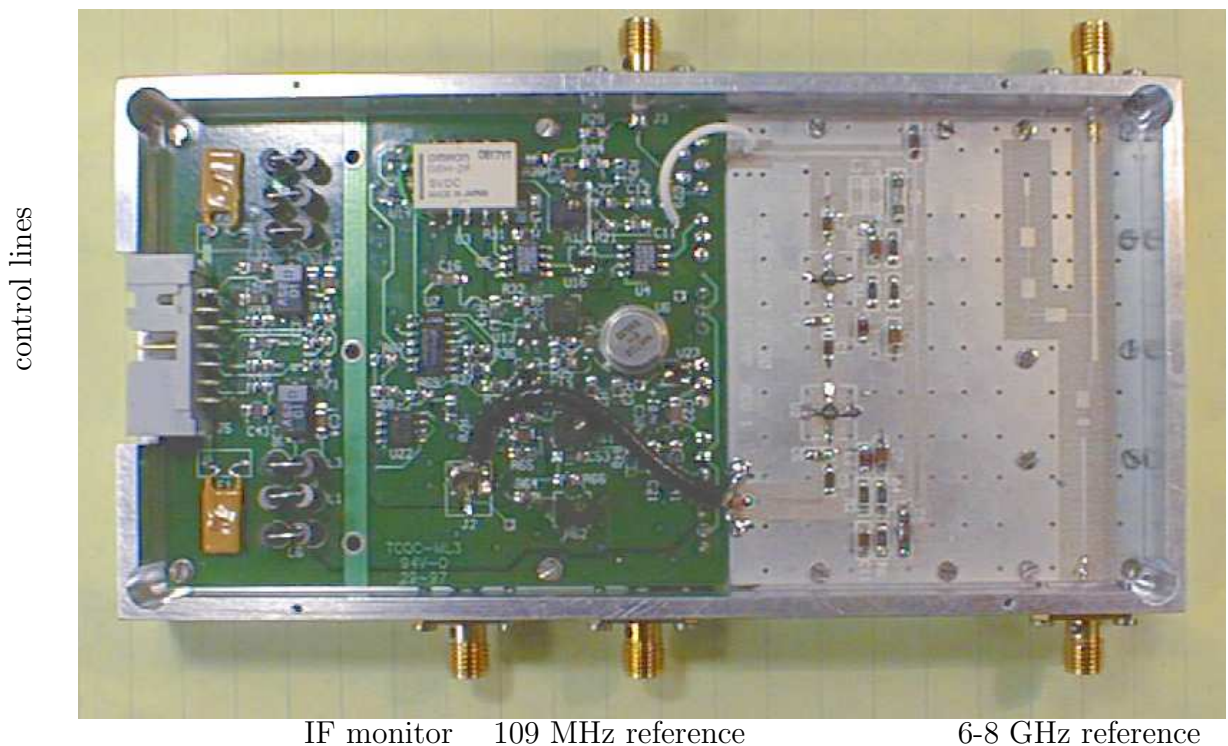,height=6.0in,angle=-90}}\\
\centerline{\hspace{1.0in} IF monitor \hspace{0.1in} 109~MHz reference \hspace{1.0in} 6-8 GHz reference}\\
\begin{rotate}{90} 
~~~~~~~~~~~~~~~~~~~~~~~~~~~~~~~~~~~~~~~~control lines
\end{rotate}
\caption{Top view of the digital PLL with the lid removed.}
\label{fig:photo}
\end{figure}

To promote the benefits of modularity and reduce the number of cables
emerging from the SMA receiver system optics cage, we decided to
provide a PLL for each LO assembly.  This requirement set constraints
on the size of the unit.  Three separate sections comprise the PLL
package shown in Figure~\ref{fig:photo}.  The righthand side of the
PLL box contains a microwave diplexer which allows the high-frequency
signal (6-8 GHz) to pass from the YIG reference input out to the
harmonic mixer on the LO chain.  At the same time, it allows the low
frequency mixing product (0-1 GHz) to return from the harmonic mixer
and to enter the two-stage IF amplifier.

\subsection{Diplexer and IF Amplifier} 

The diplexer substrate is manufactured from Arlon AR1000 Ceramic
powder-filled, woven fiberglass, PTFE composite with a dielectric
constant of 10.  The IF amplifier is comprised of two stages of
amplification via Hewlett-Packard INA-02184 Low Noise, Cascadable
Silicon Bipolar MMIC Amplifiers for a total gain of $\approx 65$~dB.
This large gain is necessary to process the tiny signal returning from
the harmonic mixer (which itself views the Gunn signal through a 20~dB
tap).  With such a large gain, suppression of noise at the amplifier
input port is crucial to prevent oscillation.  The tight-fitting lid
provides a good mechanical enclosure that prevents any noise coupling.
Also, to minimize temperature changes in the unit at power-up time
(and the accompanying phase drifts), a heater resistor is installed in
the aluminum block which can be enabled by the microcontroller when
the PLL/LO chain is turned off.


\subsection{Phase lock electronics}

\label{pllDampingPot}

A good description of PLLs is given by Gardner (1979). Our PLL is
based on a high-gain second-order loop with an active filter and a
microcontroller interface.  A schematic drawing of the circuit can be
obtained from the authors upon request.
After amplification, the IF signal is passed to the Analog Devices
AD96687 Ultrafast ECL Dual Comparator which converts the sinewave to a
squarewave.  We use zero crossing detection in the comparator in order
to avoid the need for an automatic gain control (AGC) circuit in the
IF amplifier.  Also, by biasing the comparator above the peak value of
the IF signal, the microcontroller can effectively disable the loop.
The digitized signal output from the comparator enters the Analog
Devices AD9901 Ultrahigh Speed Phase/Frequency Discriminator.  Here we
implement a loop gain control by controlling the current in the
AD9901.  Specifically, the loop gain of this system (along with the
loop bandwidth) is determined by a damping resistor (the parallel
combination of potentiometer R19 and resistor R75) and an integrating
capacitor (C9).  The choice of these values depends on the noise
characteristics of the oscillator and the loop performance
specifications.  For the SMA, the design specification requires the
PLL to reacquire lock within $t = 2\mu$s of a Walsh function phase
change.  The relationships between the rise time ($t_r$) and the 1\%
settling time of a servo loop ($t_s$) and its undamped natural
frequency ($\omega_n$) are given by:
\begin{align}
  t_r & = \frac{1.8}{\omega_n}\\
  t_s & = \frac{4.6}{\sigma} = \frac{4.6}{\zeta \omega_n}
\end{align} 
where $\zeta$ is the damping factor and $\sigma$ is the negative real
part of the pole (Franklin, Powell \& Emami-Naeini, 1994).  With a
reasonable choice of damping factor $\zeta = 0.707$, we require
$\omega_n = 2\pi \times (0.6$~MHz), which corresponds to:
\begin{equation}
  \omega_n = \sqrt{\frac{K_oK_d}{R_1C}} = 2\pi \times {\rm (0.6E+06)~radian~second}^{-1}. 
\end{equation}
where $K_o$ is the gain of the tunable oscillator in
radian~second$^{-1}$~Volt$^{-1}$, and $K_d$ is the gain of the phase
detector in Volt~radian$^{-1}$.  For Carlstrom Gunns
\citep{Carlstrom1985}, $K_o$ is on the order of 300~MHz~V$^{-1}$ =
2.0E+09 rad~s$^{-1}$~V$^{-1}$.  For our phase detector (AD9901), the
maximum value of $K_d$ is 0.285 Volt~radian$^{-1}$.  However, for
tuning flexibility, we need to operate at a gain in the middle of this
range, so we choose 0.15~V~rad$^{-1}$.  With this information in hand,
we compute the product ($R_1$C) must be 2.0E$-$05.  In the circuit, we
use $R_1$ = 1k$\Omega$ to convert the square wave into a DC
voltage. This means that C must be 20~nF.  Finally, we can relate all
of these variables back to the damping factor:
\begin{equation}
  \zeta = \frac{\tau_2\omega_n}{2} = \frac{R_2C\omega_n}{2} = 0.707.
\end{equation}
With this condition, we can solve for the final variable, $R_2$:
\begin{equation}
  R_2 = \frac{1.414}{C \omega_n} = 18\Omega.  
\end{equation}


An AD817 High Speed, Low Power Wide Supply Range Amplifier provides
the loop integrator.  A second AD817 monitors the Gunn bias voltage
and, along with the Fairchild BAV99LT1 ultrafast diode pair, keeps it
within $\pm 0.6$~V (i.e. a diode potential) of the target bias voltage
set manually via potentiometer R33.  The locking sideband can be
selected digitally via an Omron G6H-2F DC5 Low Signal Relay.  An IF
monitor port of $-20$~dB is provided for viewing the phase lock
quality on a spectrum analyzer.

\subsection{Lock detection}
By the nature of the design, phase lock is indicated whenever the gunn
bias voltage lies between the two voltage limits established by the
diode pair.  However, if the IF amplifier is being overdriven, it is
possible to acquire ``lock'' on a harmonic of the IF.  To reject these
false ``locks'', two LC circuits are included which provide a simple
yet powerful analysis of the IF to the tuning software.  A portion of
the IF signal is tapped and sent through a notch filter tuned to
109~MHz and on to an Agilent HSMS2812 Schottky Diode pair.  A similar
circuit implements a bandpass filter tuned to the same frequency.  The
voltage outputs of both filter circuits are digitized by the
microcontroller board and used in the automated phase lock search
algorithm which adjusts the tuning backshort on the Gunn oscillator
via a linear actuator while watching these voltages.

\begin{figure}[hbt] 
\centerline{\hspace{1.0in} IF monitor \hspace{0.1in} 109~MHz reference \hspace{1.0in} 6-8 GHz reference}
\centerline{\epsfig{file=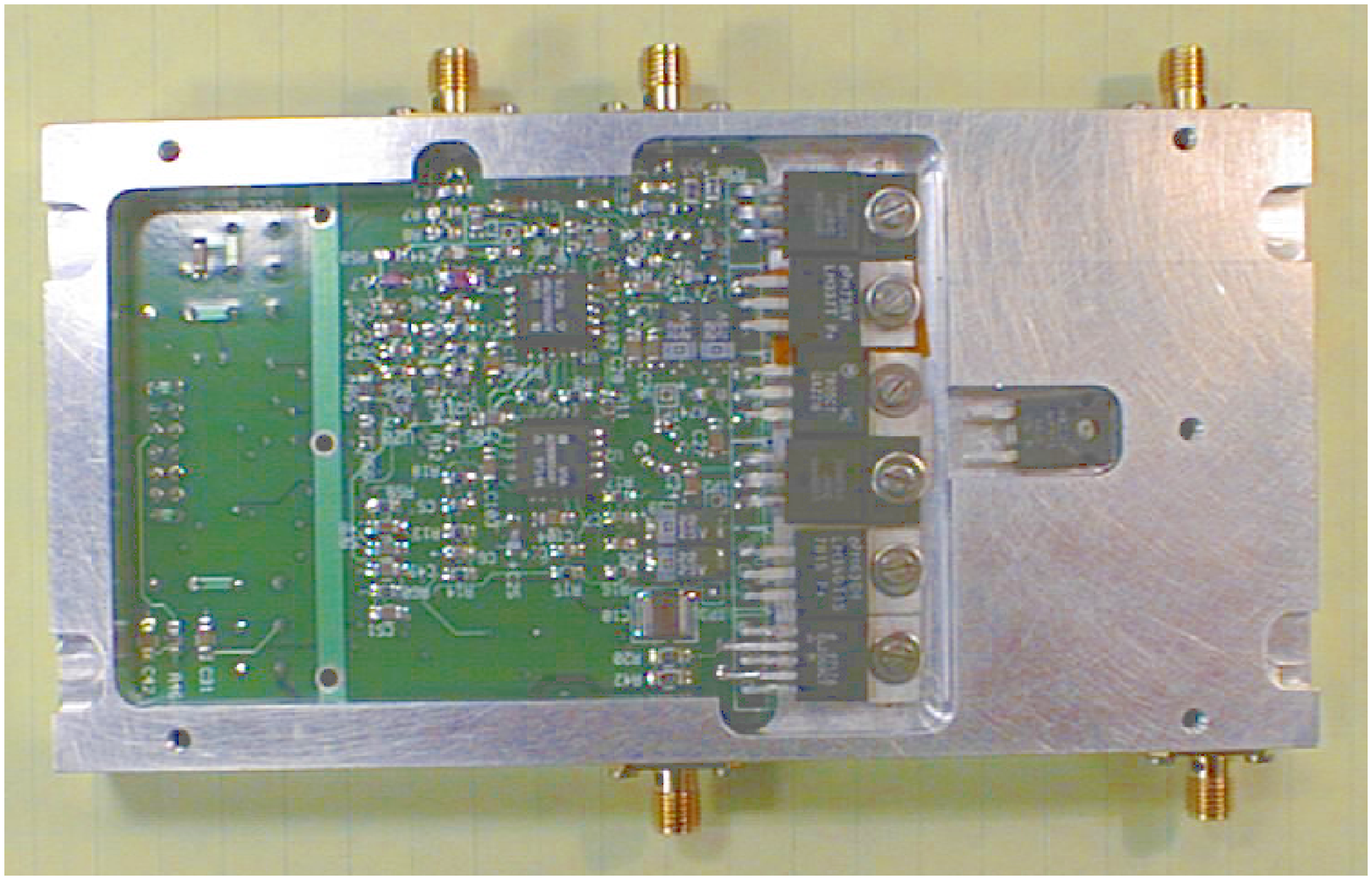,height=6.0in,angle=-90}} 
\hspace{2.5in} Gunn bias output \hspace{1.4in} Harmonic mixer\\
\caption{Bottom view of the digital PLL.}
\label{fig:photo2} 
\end{figure}

\subsection{Control interface}

The PLL is controlled through the insulation displacement connector
(IDC) port.  For manual operation, this port can be connected to a
control box with appropriate switches and an external power supply.
The remote operation of the PLLs (and their associated LO chains) is
achieved via an Intel microcontroller board connected via RS485 to the
real-time LynxOS control computer installed in a VME chassis in each
antenna.  In the laboratory setting, a standard Linux PC can be used
instead, along with an RS232/485 converter.  In either case, the
software program called ``tune6'' provides a command-line prompt and
an RPC server to implement all of the required functionality.  For
further details, see \citet{Hunter2002}.

\section{Performance}

\subsection{Adjacent channel power}

\begin{figure}[hbt] 
\centerline{\epsfig{file=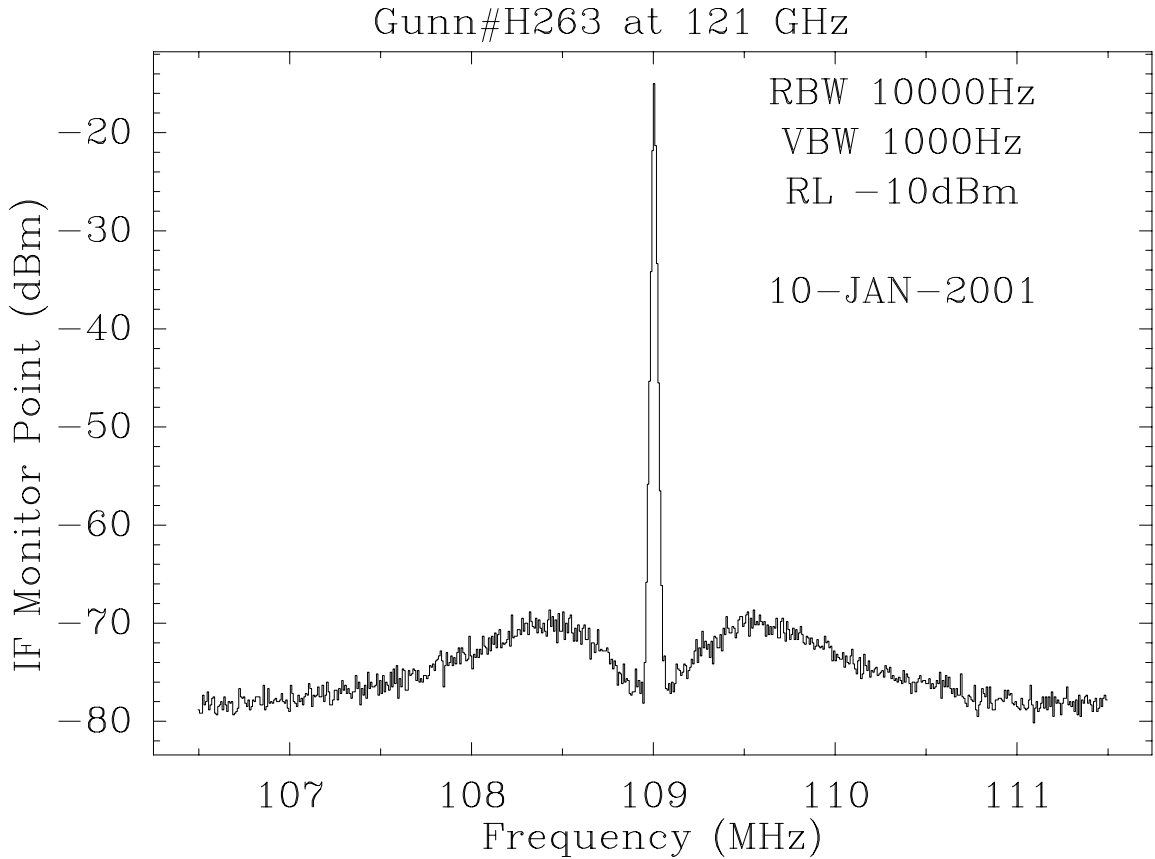,height=6.5in,angle=-90}}
\caption{Typical IF monitor signal from a Gunn oscillator phase locked 
at 121~GHz.}
\label{fig:gunn121ghz} 
\end{figure}

The quality of phase lock can be assessed by the amount of power
outside of a delta function at the IF frequency compared to the power
within the delta function.  We refer to this ratio as the adjacent
channel power and obtain values of $-35$ to $-40$~dBc on Carlstrom
Gunns using Pacific Millimeter\footnote{http://pacificmillimeter.com/}
FMA harmonic mixers (even harmonics only).  A representative plot from
a spectrum analyzer is shown in Figure~\ref{fig:gunn121ghz}.

\subsection{Walsh functions}

\begin{figure}[ht]
\centerline{\epsfig{file=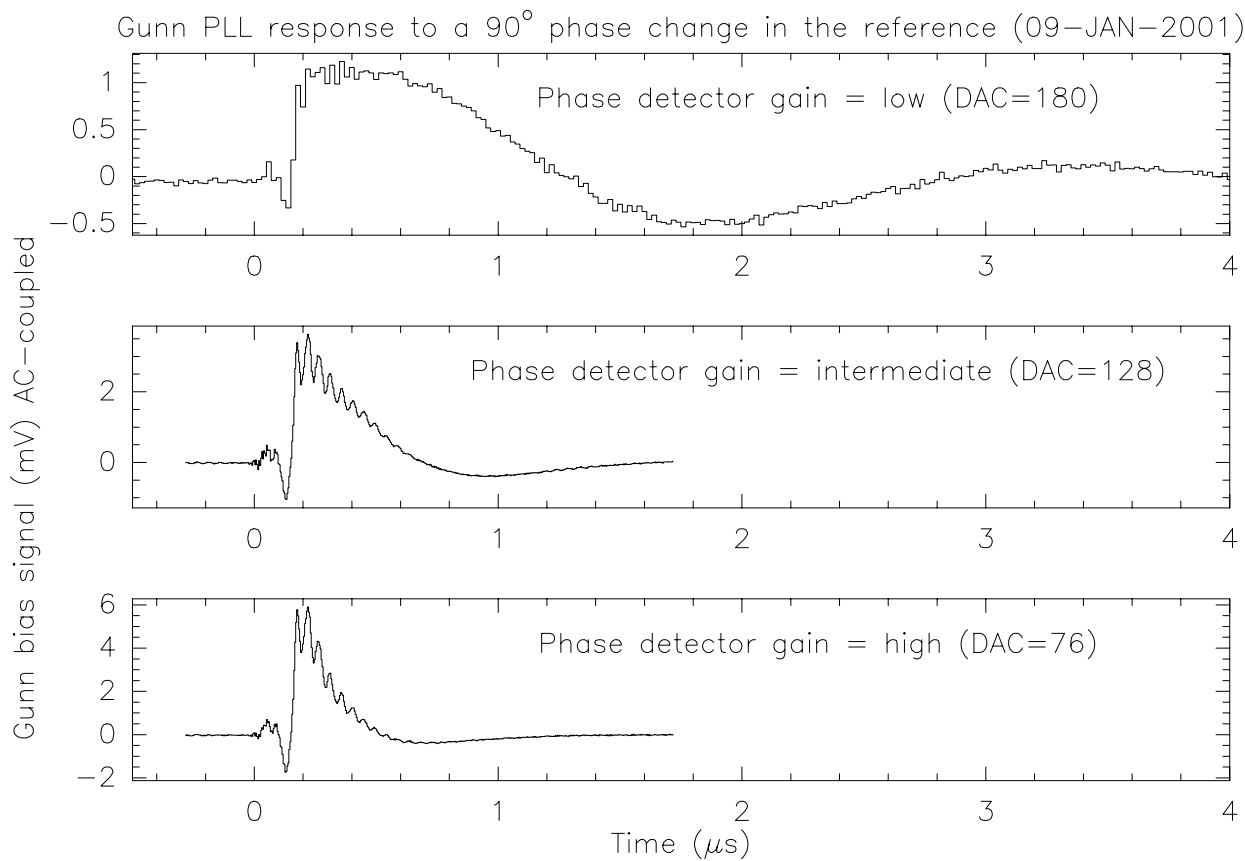,width=5.0in,angle=-90}}
\caption{Example of the fast relocking capability of the PLL for
various programmable loop gain settings.  DAC=0 corresponds to 0~Volts
and represents maximum gain (quickest response).  DAC=255 corresponds 
to 5~Volts and represents minimum gain (slowest response).}
\label{fig:relock}
\end{figure}

Walsh functions are used by radio interferometers to reject noise that
can arise in the system between the first LO and the final LO
\citep{Thompson2001}.  Our PLL easily maintains phase lock throughout
the application of phase-switching Walsh functions imposed on the
low-frequency 109~MHz reference derived from the SMA direct digital
synthesizer (DDS).  The PLL has been laboratory-tested with a
Walsh-function emulator circuit which introduces a $90^\circ$ phase
change at 100~Hz.  At the proper gain setting, it reacquires lock well
within the design specification (Figure~\ref{fig:relock}) and has
sufficient gain adjustability to compensate for the variation in the
inherent oscillator gain across the receiver band.
%

\subsection{Locking range}

The $\pm 0.6$~volt locking range provided by the PLL allows it to
maintain lock during substantial drifts of ambient temperature.  To
quantify the PLL performance, we have measured the tuning
characteristics for the two types of Gunn oscillators and present
the results in Table~1.

The main quantity of interest is the amount of temperature drift
($\Delta T$) the PLL can accommodate before losing lock. For the Gunns
tested here at this particular frequency, the corresponding values
are:
\begin{equation}
  {\rm H188}: \Delta T = (0.6{\rm V}) (150 {\rm MHz~V^{-1}}) (27.5 {\rm MHz~(^\circ C)^{-1}})^{-1} = 3.3^\circ {\rm C}
\end{equation}
\begin{equation}
  {\rm H263}: \Delta T = (0.6{\rm V}) (330 {\rm MHz~V^{-1}}) (9.8 {\rm MHz~(^\circ C)^{-1}})^{-1} = 20.2^\circ {\rm C}
\end{equation}

Clearly the Gunns with brass tuning shafts (i.e. same material as
body) are preferred for the best phase lock performance and overall
system stability.  Active control of the ambient temperature in the
antenna receiver cabin is also important.

\centerline{
\begin{tabular}{|cccccc|}
\multicolumn{6}{c}{Table 1}\\
\hline
Gunn Serial & tuning shaft & $\nu$ & $\kappa_V$ & $\kappa_L^{-1}$ & temperature coefficient\\
Number      & material     &(GHz)  & mil~GHz$^{-1}$    & MHz~Volt$^{-1}$        & MHz~($^\circ$ C)$^{-1}$\\   
\hline
H188 & stainless steel & 115.0 & 1.15 & 150 & 27.5\\
H263 & brass           & 115.0 & 1.05 & 330 &  9.8\\ 
\hline
\end{tabular}
}

\newpage
\section{Calibration of a new unit before usage}

For each new digital phase lock loop circuit that is contructed, the
following steps must be completed before it can be used on the
telescope.  Each step is described in detail below.\\
\smallskip
\noindent
$\bullet$ Place serial number label and labels on the five SMA ports\\
$\bullet$ Secure the mechanical mounting\\
$\bullet$ Set the Gunn bias potentiometer\\
$\bullet$ Set the notch filter offset potentiometer\\
$\bullet$ Set the bandpass filter offset potentiometer\\
$\bullet$ Set the notch filter center frequency\\
$\bullet$ Set the bandpass filter center frequency\\
$\bullet$ Set the PLL active loop damping potentiometer\\
$\bullet$ Put copper tape on the lid walls\\
$\bullet$ Plot the amplifier bandpass to check the gain and look for oscillations\\
$\bullet$ Confirm the microcontroller ADC value for the Gunn bias\\
$\bullet$ Measure the lock sensitivity to reference power\\
$\bullet$ Verify a clean lock trace on the spectrum analyzer\\
$\bullet$ Confirm that the loop gain is adjustable\\
$\bullet$ Measure the 109~MHz leakage\\
$\bullet$ Install the heater resistor\\

\subsection{Attach labels}

Each PLL is assigned a serial number label in the format ``dpllxx''
where ``xx'' is a decimal number with a leading zero, if necessary.
Also, the following labels should be placed on the top plate above their
respective SMA connector ports as shown in Figure~\ref{fig:photo}:
``IF MON'', ``109 REF'', ``YIG REF'', ``HARM MXR'' and ``GUNN BIAS''.

\subsection{Secure the mechanical mounting}
Be sure that the diplexer board and the digital board are securely
screwed into the box.  Failure to do so may lead to problems.

\subsection{Set the Gunn bias potentiometer}

The Gunn bias pot is R33 on the PLL board.  U7 is the LM324 14-pin
DIP.  Pin 7 of $U_7$ is OUT2 which is the point to monitor.  Measure
this voltage with respect to the chassis ground.  Adjust the pot to
bring it to the nominal voltage (e.g. 8.6 V).  You must turn on the
Gunn before doing this, of course. You may either use the
``b1testgunn'' EPROM, or use the regular ``b1t4.hex'' EPROM and issue
the ``gunn on'' command in tune6.  Now measure the Gunn bias voltage
emerging from the SMA connector.  It should be at one of the rails
(8.6$\pm$0.6 V).  If it is only 8.6$\pm$0.3 V, then you have probably
installed the resistors R36 (5k$\Omega$) and R37 (33.2k$\Omega$)
incorrectly.  Their four pads form a square.  They should be placed
horizontally on these pads, and not vertically.  If instead, the
voltage is 12 or 13V, then check the U16 diode.  It should be an
anti-parallel diode, not just a single diode.  Also, check that the
R75 resistor value is correct (two 51.1$\Omega$ resistors soldered on
top of each other).

\subsection{Set the notch filter offset potentiometer}

The notch filter offset pot is R62 on the PLL board.  Put an SMA
grounding cap on the harmonic mixer input port.  Use the ``v'' command
in tune6, or use the Palm pilot (attached to the serial bus and
running the ``busmaster'' application) to monitor the notch filter
value and adjust the pot until the value is zero or just above zero.
For reference, a single A/D bit is approximately 5 mV.  If you are
unable to set the offset to zero, then the first-stage IF amplifier
may be oscillating.  Make sure you have flattened the legs before
soldering it onto the board.

\subsection{Set the bandpass filter offset potentiometer}

The bandpass filter offset pot is R61 on the PLL board.  Follow the
same instructions as for the Notch filter offset.

\subsection{Set the notch filter center frequency}

\begin{figure}[hb]
\centerline{\epsfig{file=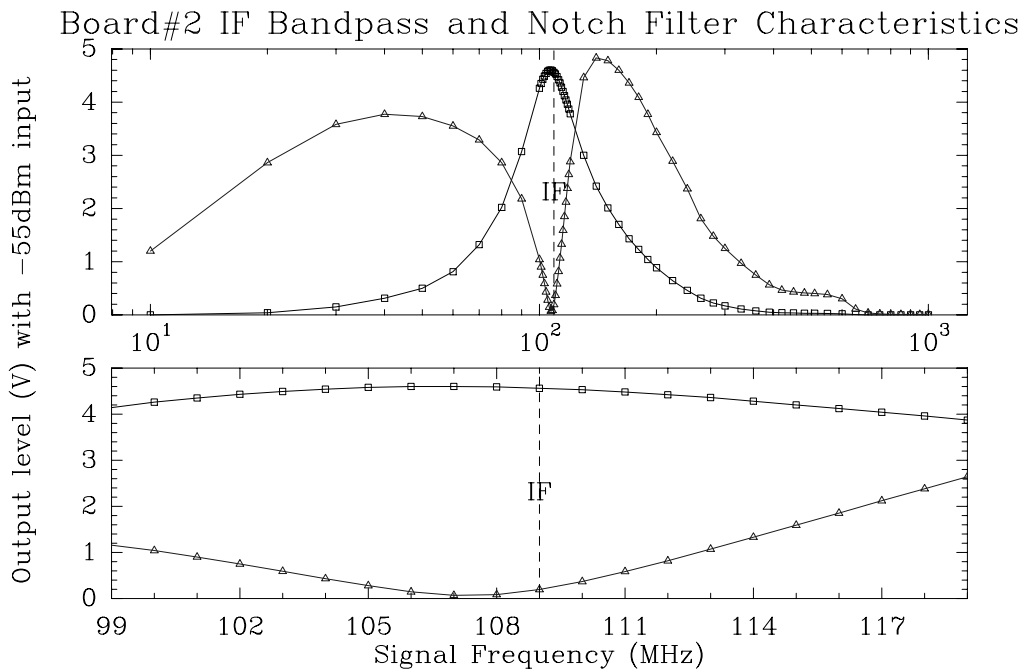,width=3.9in,angle=-90}} 
\caption{Plot of the data file generated by the tune6 command ``calibrate 
filters''.  The IF bandpass and notch filter responses are shown.}
\label{fig:iffilters}
\end{figure}

The nominal values for the filter are C44 = 4.5~pF, L8 = 470~nH and
R59 = 10~k$\Omega$.  Connect a GPIB-ready synthesizer to the harmonic
mixer input.  Run the tune6 command ``calibrate filters'' (or ``cal
fil'') to calibrate the filters.  Take the resulting data file
(if\_filters.response) and run it through the Gildas GreG macro
combinefilters.greg to plot the results (see
Figure~\ref{fig:iffilters}).  Adjust the filter component values to
bring the centers of each filter to $109 \pm 1$~MHz.  Note that a
capacitance change of 0.1~pF moves the frequency by roughly 1~MHz.
%

\subsection{Set the bandpass filter center frequency}

The nominal values for the filter are C45 = 21~pF, L7 = 100~nH and
R60 = 10~k$\Omega$.  Follow the same instructions as for the Notch
filter.

\subsection{Set the PLL active loop damping potentiometer}

The resistor $R_{2}$ is implemented by the damping pot on the PLL
board which is R19 on the silkscreen.  Measure the resistance between
the center pin and the pin above the ``1'' in ``R19'' (or,
effectively, pin 2 of U4).  Set the pot such that the resistance
equals the nominal value calculated in section~\ref{pllDampingPot}.
Try to lock a Gunn.  Once it is locked, use the tune6 ``gain'' command
to set different values of the gain (i.e. 80, 128, 160, 200).  You
should see the shoulders of the trace change position and strength.
If you see no change, then you may have a problem on the PLL board.
Check to see that R27 is not shorted out (by a solder bridge) to the
adjacent via hole.

\subsection{Place copper tape on the lid walls}
\label{copper}

To provide maximum isolation of the chamber containing the IF
amplifier, place strips of copper tape on two of the three barriers
that hang down from the box lid.  The two you want are the two closest
to each other.  On the end barrier, a single piece of 1 inch wide tape
should fold and go all the way down the barrier on both sides.  On the
middle barrier, the tape should approach no closer than 0.3 inches
from the Gunn bias edge of the box, and no closer than 0.6 inches from
the YIG reference edge of the box.  This is to avoid shorting out the
IF traces. In general, use two layers of tape on each barrier.  But be
sure that the lid does not rock.  You might need to use 3 layers on
the end barrier to prevent this.  Also, you should scrunch a bit of
copper tape into the small voids between the diplexer void and the
aluminum wall.  The bit should go under the area where the wall
touches the diplexer, so as to fully seal the chamber.

\subsection{Plot the amplifier bandpass to check the gain and look for oscillations}
\label{bandpass}

Assemble a spectrum analyzer and a synthesizer onto the GPIB bus.
Connect the spectrum analyzer to the IF monitor point of the PLL box.
With the RF power OFF, connect the synthesizer output to the harmonic
mixer input point on the PLL box.  Run the tune6 command~``testamp -s
[pllSerialNumber]'' to view and record the bandpass of the IF
amplifiers.  This command will inject $-60$~dBm signal at 109~MHz and
record a trace into the file testamp\_dpllxx.trace, and display it on
the screen using wish and a tcl script.  If the amplifiers are working
properly, you should see about $-20$~dBm at the monitor point (at 1~MHz
resolution bandwidth on the spectrum analyzer).  Now set the frequency
to 800~MHz.  You should see the signal reduced to about $-55$~dBm at the 
monitor point.

\begin{figure}[htb]
\centerline{\epsfig{file=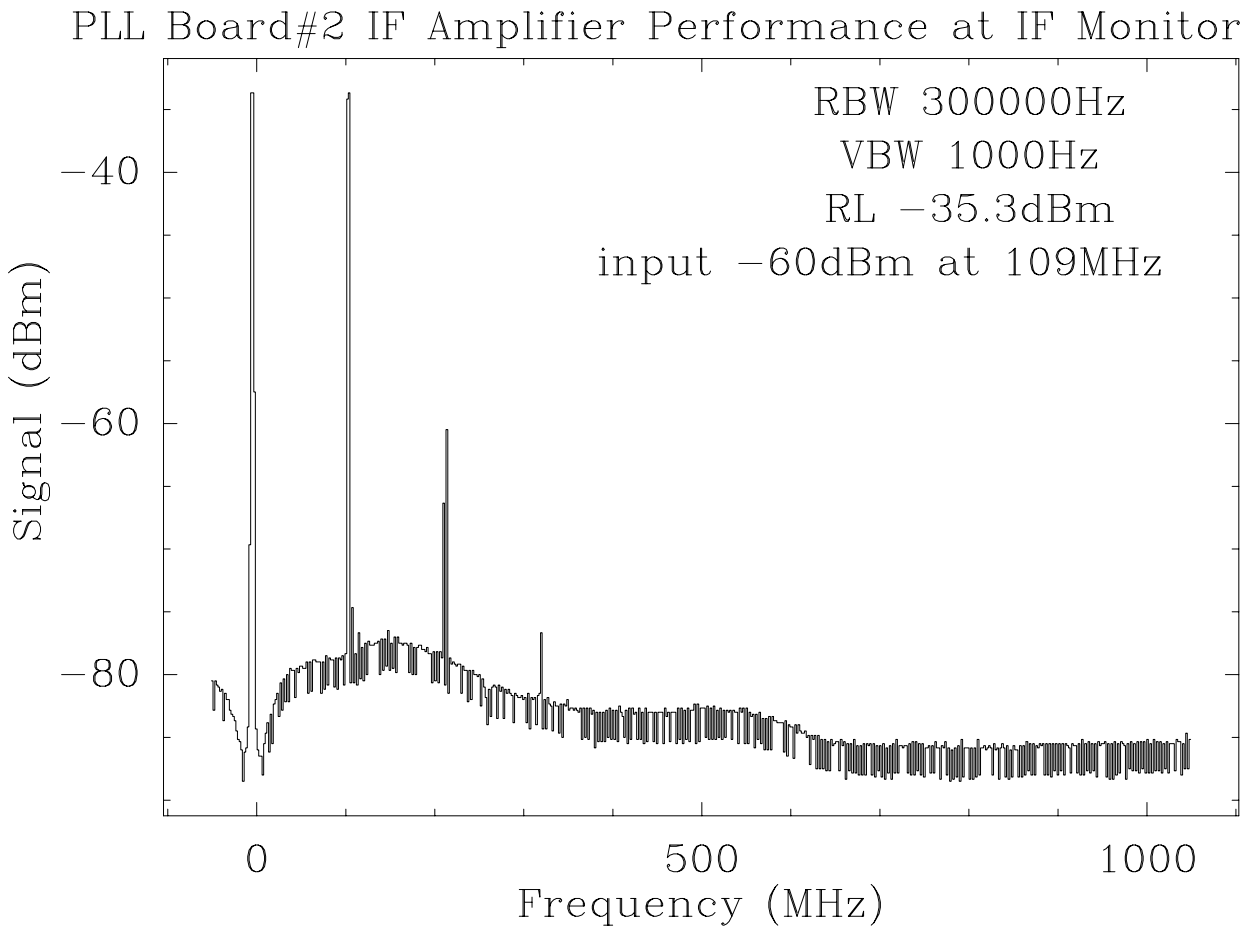,width=4.9in,angle=-90}}  
\caption{Plot of the data file generated by the tune6 command ``testamp''.
The IF amplifier bandpass is shown along with the test tone (and its 
harmonics).}
\label{fig:amplifiers}
\end{figure}

Now, disconnect everything from the harmonic mixer port.  The response
curve should remain smooth with no oscillations (see
Figure~\ref{fig:amplifiers}).  If instead you see a strong signal
(anything above $-50$~dBm) at 700 or 800~MHz, it is an IF amplifier
oscillation that should be fixed.  To fix it, first remove the PLL box
lid and verify that bandwidth control capacitor (C4) is a porcelain
33~pF device (made by ATC).  Second, verify that inductor L2 is 220nH.
Finally, examine the solder joints of the two IF amplifiers.  Solder
should be placed on all the legs as close as possible to the device.
This should kill the oscillation.  If not, then revisit the section on
copper tape (section~\ref{copper}), with particular attention to the
voids between the diplexer board and the wall of the box.

If you do not want to open the PLL box to fix it, a stop-gap measure
is to place a 1~dB pad on the output to the harmonic mixer.  This will
also kill the oscillation.  Even if you do nothing, once the
$\sim$10dBm synthesizer (or YIG) signal is applied, the oscillation
always disappears (since the impedance of the harmonic mixer changes).

\subsection{Confirm the microcontroller ADC value for the Gunn bias}

Place the Gunn bias output on a multimeter.  Run ``v'' in tune6 (or
''b1testADCs'' on the Palm) and look at the digitized Gunn bias value.
The two values should be within 5-10 mV.  If they are not, first check
that the voltage-dividing resistor network has the correct values
(20k$\Omega$ for R34 and 10k$\Omega$ for R35), and that the other ADC
inputs are not saturated.  If the resistor values are correct, then
you should add trimming resistors on top of the original resistors
to bring the digitized voltage into better agreement with the actual voltage.
Use the following formulas:
\begin{equation}
Q = \frac{V_{ADC}}{V_{true}}
\end{equation}
\begin{equation}
{\rm if~} V_{ADC} > V_{true}, {\rm then~add~to~R35:~} R= \frac{20000}{3(Q-1)}\\
\end{equation}
\begin{equation}
{\rm if~} V_{ADC} < V_{true}, {\rm then~add~to~R34:~} R = \frac{20000}{3}\frac{(3Q-2)}{(1-Q)}\\
\end{equation}

\subsection{Measure the lock sensitivity to reference power}

First lock the Gunn to some frequency.  Then turn down the 109~MHz
reference signal power to $-30$~dBm and see if phase lock is
maintained.  This can be accomplished by installing a 20~dB and a
10~dB attenuator in line with the Fluke 6160B synthesizer set to its
lowest output power setting.  A PLL in good shape will be able to
retain lock.  Next, disable both references, the Gunn itself and the
PLL for a time, then re-enable then.  The PLL should re-acquire lock
even at this low reference level.  The power level where lock is lost
appears to be $-35$~dBm, while lock is regained at $-33$~dBm.

\subsection{Verify a clean lock trace on the spectrum analyzer}

The lock signal should look like Figure~\ref{fig:gunn121ghz}, with
symmetric shoulders surrounding the delta function.  If there are
sidebands around 860~kHz and 1700~kHz offset from the signal, then
check to see if the bypass capacitors (e.g. C30) around the negative
regulators are present.  Without these, the regulators can oscillate
and produce artifacts in the lock loop.  In a wider frequency span
display, you might also see faint lines at 0.5 and 1.5 times the IF
frequency.  We believe these are due to the fundamental frequency of
the Gunn (0.5 times its primary output frequency) propagating to the
harmonic mixer and mixing with half the harmonic of the YIG reference
compared to the main Gunn signal.  The evidence for this is that the
lines disappear if you tune the YIG to an odd harmonic (since the
fundamental cannot mix with a non-integer harmonic).  In any case,
these lines should not be anything to worry about.

\subsection{Confirm that the loop gain is adjustable}

Using the tune6 command ``gain'', change the loop gain within the
range from 0 to 255.  You should see a corresponding change in the
appearance of the lock signal in the IF spectrum.  Lower values
correspond to lower DAC voltages which correspond to higher loop gain,
which should spread the shoulders out to higher frequencies.  If there
is no noticeable change in the trace, then there may be a problem on
the PLL circuit.  The pads for R27 sit close to an adjacent ``via'' on
the circuit board which can be shorted to if a blob of solder
overflows the pad.  This error will prevent the gain from changing.

\subsection{Measure the 109 MHz leakage}

Some of the 109~MHz reference power entering the PLL box can leak back
into the IF amplifier and mimic the signal returned from the harmonic
mixer.  At low levels ($<-40$~dBm as seen on the IF monitor port) this
leakage is not a problem.  However, larger values (e.g. $-33$~dBm) can
prevent the PLL from locking as it rivals the power of the mixed-down
Gunn signal.  Connect a 109~MHz signal of +13~dBm (this is the maximum
that the Fluke 6160B synthesizer can deliver) into the 109~MHz
reference port on the PLL and look at the IF monitor signal on the
spectrum analyzer.  If the value is above $-40$~dBm, try putting down
more copper tape (as in section~\ref{copper}) to reduce the level.

\subsection{Install the heater resistor}

One side of the $50\Omega$ heater resistor should be connected to the
test point TP1 (which goes to the collector = pin 3 of U23).  The
other side of the resistor should be connected to ground (we solder it
to pin 1 of the U11 regulator).  The emitter of U23 is connected to
+5V, thus when the heater is enabled, there should be only 100mA
drawn.  The collector pin is labelled ``FETBIAS'' on the schematic.
If you see 18V across the resistor, then somehow you have wired it
incorrectly.

\subsection{What if it doesn't lock?}

\begin{enumerate}
\item{
If everything is set right and the PLL refuses to lock, you should
first check the quality of the soldering on the SMA connectors under a
microscope.  The solder joint on the SMA connector pin carrying the
109~MHz reference may need to be reheated.}
\item{  If the solder connections are
okay, then check out the phase-frequency detector.  Put an RF signal
(30-150~MHz) on the reference and signal inputs (with some phase shift
between them).  You can do this by connecting a synthesizer output to
the 109~MHz reference input port on the PLL, and through 60~dB of
attenuation to the harmonic mixer port.  Look at the output of the
comparator (U1) on an oscilloscope.  You should see an RF signal.
Then compare the input and output pins of the AD9901 phase-frequency
discriminator (U2).  If you don't see any output, try changing the
chip.  You should see the phase of the two waveforms change as you
lower the frequency of the signal.  Be sure to blink the PLL everytime
you change frequency or else the chip may go into never-never land and
you won't see a signal. If after blinking it, you still don't see the
proper output, you may have to replace the AD9901.}
\end{enumerate}

\subsection{What if it oscillates at 700 or 800 MHz?} 

See section~\ref{bandpass}.

\subsection{Confirm the lock indication on J3 motor connector} 

The lock status is also sent to pins 6 (+) and 1 (-) on the J3 motor 
connector.  The signal is +12V.  This may be used at other telescopes 
(such as CSO and JCMT) for an external lock LED indicator (through
a dropping resistor).

\acknowledgments We thank the Smithsonian Institution for supporting
this project and Irwin I. Shapiro for his encouragement. This research
has made use of NASA's Astrophysics Data System Bibliographic
Services.  We extend special thanks to those of Hawaiian ancestry on
whose sacred mountain we are often privileged to be guests.


\begin{thebibliography}{}

\bibitem[Blundell(2005)]{Blundell2005} Blundell, R.\ 2005, 
arXiv:astro-ph/0508492 

\bibitem[Bottinelli et al.(2008)]{Bottinelli2008} Bottinelli, S., et 
al.\ 2008, \procspie, 7012,  

\bibitem[Carlstrom et al.(1985)]{Carlstrom1985} Carlstrom, J.~E., 
Plambeck, R.~L., 
\& Thornton, D.~D.\ 1985, IEEE Transactions on Microwave Theory Techniques, 33, 610 

\bibitem[Doeleman et al.(2008)]{Doeleman2008} Doeleman, S.~S., et 
al.\ 2008, \nat, 455, 78 

\bibitem[Doriese(2002)]{Doriese2002} Doriese, W.~B.\ 2002, 
Ph.D.~Thesis, Princeton University

\bibitem[Fowler et al.(2005)]{Fowler2005} Fowler, J.~W., et al.\ 
2005, \apjs, 156, 1 

\bibitem[Franklin, Powell \& Emami-Naeini(1994)]{Franklin1994} 
Franklin, G.F., Powell, J.D. \& Emami-Naeini, A., \underline{Feedback
Control of Dynamic Systems}, Addison-Wesley, Reading, MA: 1994

\bibitem[Gardner(1987)]{Gardner1987} Gardner, F.M,
  \underline{Phaselock Techniques}, Wiley \& Sons, New York, NY:
  1987

\bibitem[Hunter et al.(1996)]{Hunter1996} Hunter, T.~R., Benford, 
D.~J., \& Serabyn, E.\ 1996, \pasp, 108, 1042 

\bibitem[Hunter et al.(2002)]{Hunter2002} Hunter, T.~R., Wilson, 
R.~W., Kimberk, R.~S., Leiker, P.~S., 
\& Christensen, R.~D.\ 2002, \procspie, 4848, 206 

\bibitem[Hunter et al.(2006)]{Hunter2006} Hunter, T.~R., Zhao, 
J.~H., Liu, S.-Y., 
\& Su, Y.-N.\ 2006, USNC-URSI National Radio Science Meeting at University of Colorado, January 4-7, 2006, 

\bibitem[Hunter et al.(2007)]{Hunter2007}Hunter, T.~R., et al.\ 
2007, arXiv:0704.2641 

\bibitem[Hunter et al.(2008)]{Hunter2008} Hunter, T.~R., Brogan, 
C.~L., Indebetouw, R., \& Cyganowski, C.~J.\ 2008, \apj, 680, 1271 

\bibitem[Kawamura et al.(1999)]{Kawamura1999} Kawamura, J.~H., 
Hunter, T.~R., Tong, C.-Y.~E., Blundell, R., Zhang, Q., Katz, C.~A., Papa, 
D.~C., \& Sridharan, T.~K.\ 1999, \pasp, 111, 1088 

\bibitem[Kimberk et al.(2006)]{Kimberk2006} Kimberk, R., Hunter, 
T., Tong, C.-Y.~E., 
\& Blundell, R.\ 2006, Seventeenth International Symposium on Space Terahertz Technology, 206 

\bibitem[Moran(1996)]{Moran1996} Moran, J.~M.\ 1996, Bulletin of 
the American Astronomical Society, 28, 895 

\bibitem[Thompson et al.(2001)]{Thompson2001} Thompson, A.~R., 
Moran, J.~M., 
\& Swenson, G.~W., Jr.\ 2001, Interferometry and synthesis in radio astronomy by A.~Richard Thompson, James M.~Moran, and George W.~Swenson, Jr.~2nd ed.~ New York : Wiley, c2001.xxiii, 692 p.~: ill.~; 25 cm.~''A Wiley-Interscience publication.''  Includes bibliographical references and indexes.~ISBN :  0471254924,  

\bibitem[Torbet et al.(1999)]{Torbet1999} Torbet, E., et al.\ 
1999, \apjl, 521, L79 

\bibitem[Weintroub(2008)]{Weintroub2008} Weintroub, J.\ 2008, Journal 
of Physics Conference Series, 131, 012047 

\end{thebibliography}
\end{document}